\documentclass[FBSedit,FBSmath,ecsub]{FBSsuppl}
\usepackage{amsfonts}
\usepackage{amssymb}
\newif\ifpdf
	\ifx\pdfoutput\undefined
	\pdffalse 
	\else
	\pdfoutput=1 
	\pdftrue
	\fi
\ifpdf
	\usepackage[pdftex]{graphicx}
	\DeclareGraphicsExtensions{.pdf}
\else
	\usepackage{graphicx}
	\DeclareGraphicsExtensions{.eps}
\fi

\title{Borromean molecules%
\footnote{Talk given at the XVIII European Few-Body Conference, Bled, Slovenia, September 2002, to app. in Few-Body Sys. Suppl.\ preprint ISN 02-73; Archive physics/0210018}}
\author{Jean-Marc Richard\thanks{\textit{E-mail address:} 
jean-marc.richard@isn.in2p3.fr}}
\institute{Institut des Sciences Nucl\'eaires, Universit\'e Joseph 
Fourier -- CNRS-IN2P3, 
53, avenue des Martyrs, F-38026 Grenoble Cedex, France}
\begin{document}
\maketitle
\begin{abstract}
For some values of the constituent masses $m_i$, the hydrogen-like molecule 
$(m_1^+,m_2^+,m_3^-,m_4^-)$ is stable with respect to spontaneous dissociation into two neutral atoms, while none of its three-body subsystems such as 
$(m_1^+,m_2^+,m_3^-)$ is stable. This occurs in particular in the neighbourhood of
the $(M^+,m^+,M^-,m^-)$ configuration with $M/m=2$, as for instance for the molecule
$(p,d,\bar{p},\bar{d})$ involving an antiproton and an antideuteron.
\end{abstract}

The word ``Borromean'' is often used to identify bound states whose subsystems are unbound \cite{Bang93}. It comes after the Borromean rings, which are interlaced in such a subtle topological way, that if any one of them is removed, the two others become unlocked. For instance, the ${}^6\mathrm{He}$ isomer of ordinary Helium is stable, while ${}^5\mathrm{He}$ is not. In a three-body picture, this means that the ($\alpha, n, n)$ system is bound, while neither $(\alpha, n)$ nor $(n,n)$ are stable against dissociation.

For short-range forces, Borromean binding occurs explicitly in simple quantum-mechanical models. 
Let $g$ be the strength of the pair potential $V$ acting between $N$ identical particles and  containing attractive parts. Binding two bosons requires a certain minimum strength, say $g>g_2$. For three bosons, this is $g>g_3$. The remarkable observation is that $g_3<g_2$. Typically, $g_3\simeq0.8\,g_2$ for simple monotonic potentials such as Yukawa, Gaussian, exponential, etc. See, e.g., \cite{Bang93}, and references therein. It can be proved that $g_3/g_2\ge 2/3$ \cite{Bang93}. This lower limit is approached for potentials with an external barrier, whereas $g_3/g_2\to 1$ for those with a strong repulsive core \cite{Bang93}.
Similarly $g_4<g_3$, etc. A 4-body system is Borromean for $g_4< g< g_2$, if one adopts the following definition: {\sl A  bound state is Borromean if there is no path to build it via a	 series of  stable states by adding the constituents one by one.}

The phenomenon of Borromean binding is intimately linked to the Thomas collapse \cite{Thomas35} and the Efimov effect \cite{Efimov70}. Let $E_N$ be the ground-state energy of $H_N$. For $g\to g_2+$, $E_2/E_3\to 0$, thus if $E_2$ is kept fixed by rescaling, $E_3\to -\infty$. The occurrence of many excited states  near $g=g_2$ implies that the ground state   already exists at this point.

Borromean binding is hardly conceivable for long-range attractive potentials. For  the gravitational $V(r)\propto-1/r$ potential, the critical strength is $g_N=0$, $\forall N$, by scaling considerations. The situation is less obvious for a superposition of attractive and repulsive terms,  and different masses. Still the existence of a bound state survives any overall rescaling of the strength factors  or  masses, but depends rather crucially on the balance between the repulsive and attractive terms, and on the mass ratios.

\begin{figure}
\begin{minipage}{.45\textwidth}
\centering{\includegraphics[width=.8\textwidth]{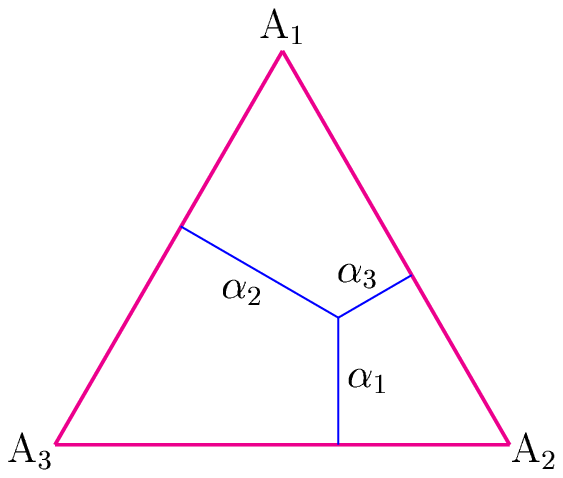} }
\end{minipage}
\begin{minipage}{.45\textwidth}
\centering{\includegraphics[width=.8\textwidth]{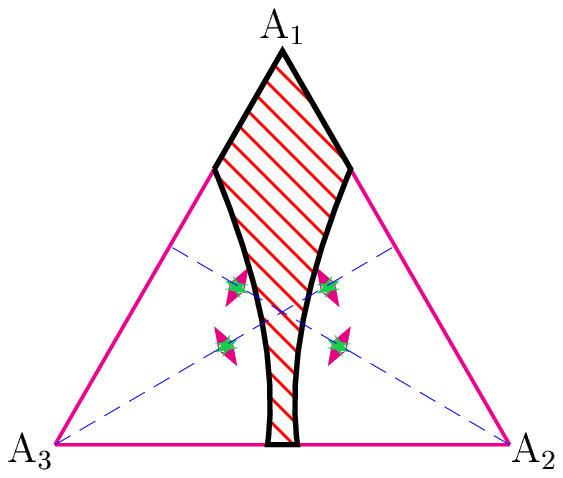}  }
\end{minipage}
\caption{Geometrical representation of the stability domain for  3-body systems $(m_1^+,m_2^-,m_3^-)$ represented by their normalised inverse masses $\alpha_i$.
The arrows indicate the range explored by the 3-body subsystems of
$[(A\pm\epsilon)^+,a^+,(A\mp\epsilon)^-,a^-]$ and $[A^+(a\pm\eta)^+,A^-(a\mp\eta)^-]$, for
inverse masses in ratio $a/A=2$. Here $\epsilon=0.1$ and $\eta=0.25$, but the stability border is drawn without accurate computation.\label{Fig:Tri}}
\end{figure}

The systems of three unit charges have been studied extensively. See, e.g., \cite{Armour93} for a comprehensive review. While He$^+(e^-,p,p)$, Ps$^-(e^+,e^-,e^-)$, and H$^-(p,e^-,e^-)$, for instance, are stable against dissociation into a neutral atom and an isolated  charge, configurations such as $(e^-,e^+,p)$ or $(p,\bar{p},e^-)$ are unstable. Only coulomb forces are considered here, and finite-size effects, strong interaction, annihilation, etc., are neglected. The stability frontier is shown in Fig.~\ref{Fig:Tri}. Thanks to scaling, any state is represented by its normalised inverse masses $\alpha_i=m_i^{-1}/\sum_k m_k^{-1}$. The domain $\alpha_1+\alpha_2+\alpha_3=1$, $\alpha_i>0$, is limited by an equilateral triangle, see Fig.~\ref{Fig:Tri}. There is a stability band around the symmetry axis, in between two  instability domains which are convex \cite{Armour93}.

The  $(m^\mp,M^\pm,m^\pm)$ configurations have been studied in some detail. They correspond to the oblique symmetry axes  in Fig.~\ref{Fig:Tri}. For $m=M$, one gets the stable Ps$^-$. For $M\gg m$, $(e^-,p,e^+)$ is unstable. For $M\ll  m$, $(p,e^-,\bar{p})$ is also unstable. The stability range has been estimated by Mitroy \cite{Mitroy2000}, 
\begin{equation}\label{Mitroy}
0.70\lesssim M/m \lesssim1.63~,
\end{equation}
using a powerful and  reliable stochastic variational method \cite{Varga95}. 

There are fewer rigorous or empirical results on $(m_1^+,m_2^+,m_3^-,m_4^-)$ systems  with charges $\{q_i\}=\{\pm1,\pm1,\mp1,\mp1\}$.
Still, valuable information exists in the literature \cite{Armour93}. The configuration $(p,e^+,\bar{p},e^-)$ is unstable, as it cannot compete with the deeply bound protonium associated with the lowest threshold $(p,\bar{p})+(e^+,e^-)$. On the other hand, the hydrogen molecule H$_2(p,p,e^-,e^-)$, the positronium molecule  Ps$_2(e^+,e^+,e^-,e^-)$, the positronium hydride PsH$(p,e^+,e^-,e^-)$ are known to be stable. Any configuration with two identical particles, e.g., $m_3=m_4$, are stable against spontaneous dissociation. In particular, any hydrogen-like state $(M^+,M^+,m^-,m^-)$ is stable with respect to its lowest threshold $2(M^+,m^-)$, with stability improved if $M/m$  is increased or decreased, starting from the Ps$_2$ case $M/m=1$ \cite{Armour93}

The above examples share a common property: Ps$_2$ contains the $(e^\pm,e^\mp,e^\mp)$ subsystem, which possesses a stable bound stable; for H$_2$, both $(e^-,p,p)$ and $(p,e^-,e^-)$ have at least one stable bound state.
For PsH, $(e^-,e^+,p)$ is unstable, but $(p,e^-,e^-)$ and $(e^+,e^-,e^-)$ have a stable bound state. Any $(m_1^+,m_2^+,m^-,m^-)$ state is stable, with at least $(m_1^+,m^-,m^-)$ and $(m_2^+,m^-,m^-)$ having a stable bound state. It is tempting at this point to conjecture that any stable molecule has at least one stable 3-body subsystem, i.e., can be build step by step with stable intermediate states: single charge, atom, and  3-body ion.

This conjecture is wrong, as can be seen by examining the second set of $N=4$ configurations with two masses. This is $(M^+,m^+,M^-,m^-)$, which is invariant under charge conjugation. For $M\gg m$ or $M\ll m$, it corresponds to the unstable $(p,e^+,\bar{p},e^-)$, while for $M=m$, it leads to a rescaled Ps$_2$ which is stable. The critical mass ratio at which instability occurs has been estimated by  accurate  methods  \cite{Bressanini97} 
\begin{equation}\label{MmMm}
{1\over 2.2}\lesssim {M\over m}\lesssim2.2~.
\end{equation}
Comparing the results (\ref{Mitroy}) and (\ref{MmMm}) indicates a window for Borromean binding. In particular, for $M/m=2$, which corresponds for instance to a deuteron and a proton, the molecule $(M^+,m^+,M^-,m^-)$ is bound, but neither $(M^\pm,m^\mp,M^\pm)$ nor $(m^\pm,m^\mp,M^\pm)$ are stable. The molecule experiences a new type of binding mechanism, which is more fragile.
It would be interesting to investigate the wave function in some detail, and study to what extent it differs from that of the positronium molecule.

One could also resume the numerical investigations of Ref.~\cite{Bressanini97} and determine the position of stability border of $(m_1^+,m_2^+,m_3^-,m_4^-)$ in the vicinity  of the $(m_1=m_3,\,m_2=m_4)$ line. Already, some minimal extension of the domain of Borromean binding can be made without any new 4-body calculation, provided the stability border is accurately known in the 3-body case. Consider for instance the configuration given by the {\em inverse} masses 
\begin{equation}
\label{exploration}
[(A\pm\epsilon)^+,a^+,(A\mp\epsilon)^-,a^-]~,\ \hbox{or}\ \ 
[A^+(a\pm\eta)^+,A^-(a\mp\eta)^-]~,
\end{equation}
where $A=M^{-1}$ and $a=m^{-1}$ links to the previous notation, for $\epsilon=\eta=0$.
If $\vert\epsilon\vert<\vert a-A\vert$, the lowest threshold remains given by
$[(A\pm\epsilon)^+,(A\mp\epsilon)^-]+[a^+,a^-]$, which has the same energy as for $\epsilon=0$, since the inverse masses average out when building the reduced mass for each atom. On the other hand, the four-body Hamiltonian for the first configurations can be written
\begin{equation}
\label{epsilon}
H(\epsilon)=H(0)\pm{\epsilon\over2}(\mathbf{p}_1^2-\mathbf{p}_3^2)~.
\end{equation}
The second term breaks charge-conjugation symmetry and lowers the ground state. This can be seen by applying the variational principle with the symmetric ground-state wave-function of $H(0)$ as a trial wave-function for $H(\epsilon)$. Hence, the 4-body molecule has an improved stability for $\epsilon\neq 0$ in (\ref{epsilon}). As $\epsilon$ departs from 0, the 3-body subsystems follow transverse and longitudinal paths starting from the $\alpha_1=\alpha_3$ or $\alpha_1=\alpha_2$ symmetry axis of the triangle, as pictured schematically in Fig.~\ref{Fig:Tri}. It should be not too difficult to determine up to which values of $\epsilon$ the 3-body systems remain unstable.
The same reasoning can be applied to the second configurations.

For example,  if the mass ratio of the  protonium molecule $(p,p,\bar{p},\bar{p})$ is modified by replacing a proton and an antiproton by a deuteron and an antideuteron, the resulting molecule  $(p,d,\bar{p},\bar{d})$  becomes Borromean, without any bound 3-body subsystem. This can be compared with the observation by Bertini et al.\ that if one screens the Coulomb potential in the hydrogen molecule, that is to say, $r_{ij}^{-1}\to\exp(-\mu r_{ij})/r_{ij}$ for all pairs, there is a range of values of $\mu$ for which H$_2$ remains bound while neither H$_2^+$ nor  H are stable \cite{Bertini2002}.

This investigation grew out of discussions with Dario Bressanini and Andr\'e Martin. Comments by E.A.G.~Armour, A.S.~Jensen and K.~Varga  are also gratefully acknowledged. I benefited from the stimulating atmosphere of the XVIII European Few-Body Conference at Bled, where this work was presented and discussed with my colleagues.


\small
\begin{thebibliography}{10}

\bibitem{Bang93}
M.V. Zhukov, B.V. Danilin, D.V. Fedorov, J.M. Bang, I.S. Thompson, and J.S.
  Vaagen, Phys. Rep. {\bf 231} (1993) 151;
J.-M. Richard and S. Fleck, Phys. Rev. Lett. {\bf 73} (1994) 1464;
S. Moszkowski, S. Fleck, A. Krikeb, L. Theu\ss l, J.-M. Richard, and K. Varga,
  Phys. Rev. A 62 (2000) 032504.

\bibitem{Thomas35}
L.H. Thomas, Phys. Rev. {\bf 47} (1935) 903; A. Delfino, K. Adhikari, L. Tomio
  and T. Frederico, Phys. Rev. {\bf C46} (1992) 471.

\bibitem{Efimov70}
V. Efimov, Phys. Lett. {\bf 33B} (1970) 560; Sov. J. Nucl. Phys. {\bf 12}
  (1971) 581; R.N. Amado and R.V. Noble, Phys. Lett. {\bf 35B} (1971) 25; Phys.
  Rev. D {\bf 5} (1972) 1992; S. Albeverio, R. H{\o}egh-Krohn and T.T. Wu,
  Phys. Lett. {\bf 83A} (1981) 101; D.V. Fedorov and A.S. Jensen, Phys. Rev.
  Lett. {\bf 71} (1993) 4103.

\bibitem{Armour93}
E.A.G. Armour and W. Byers Brown, Accounts of Chemical Research, {\bf 26}
  (1993) 168; E.A.G. Armour, J.-M. Richard, and K. Varga, ``Stability of Few-Charge
  Systems'', in preparation.

\bibitem{Mitroy2000}
J.~Mitroy, J. Phys. B: At. Mol. Opt. Phys. {\bf 33} (2000) 5307.

\bibitem{Bressanini97}
D. Bressanini, M. Mella and G. Morosi, Phys. Rev. A {\bf 55} (1997) 200;
K. Varga, Few-Body Systems Suppl.\ {\bf 10} (1999) 11, (Proc. 16$^\mathrm{th}$
  European Conference on Few-Body Problems in Physics, Autrans, France, June
  1998, ed.\ B.~Desplanques et al.).

\bibitem{Bertini2002}
L.~Bertini, M.~Mella, D.~Bressanini, and G.~Morosi, ``A non-adiabatic quantum
  Monte-Carlo study of the critical stability of H$_2^+$ and H$_2$ with Yukawa
  potential'', preprint (2002).

\end{thebibliography}
\end{document}